# SURVIVAL OF *SALMONELLA* SPP. IN COMPOSTING USING VIAL AND DIRECT INOCULUMS TECHNIQUE

SUNAR, N. M.*+, STENTIFORD, E.I.*, FLETCHER, L.A* AND D.I.STEWART*

+ *Department of Environmental Engineering, Faculty of Engineering, University of Tun Hussien Onn Malaysia, Batu Pahat, Johor, Malaysia.*
\**Pathogen Control Engineering (PaCE) Institute*, *School of Civil Engineering, University of Leeds, Leeds, LS2 9JT, United Kingdom.*
+**Corresponding Author**. *Tel: +44 (0) 113 3432319. Email: shuhaila@uthm.edu.my*

## EXECUTIVE SUMMARY

The survival of *Salmonella* spp. as pathogen indicator in composting was studied. The inoculums technique was used to gives the known amounts of *Salmonella* spp. involved in composting. The inoculums of *Salmonella* spp. solution was added directly into the compost material. The direct inoculum was compared with inoculums in vial technique. The *Salmonella* spp. solution placed into a vial and inserted into the middle of compost material before starting the composting process. The conventional method that is used for the enumeration of *Salmonella* spp. is serial dilution followed by standard membrane filtration as recommended in the compost quality standard method PAS 100 and the British Standard (BS EN ISO 6579:2002). This study was designed to investigate the relationship of temperature and contact material that may also involve in pathogen activation specifically to *Salmonella* spp. The exposure to an average temperature during composting of about 55-60°C was kept for at least 3 days as it was reported sufficiently kills the vast majority of enteric pathogen (Deportes et al., 1995). The amount of *Salmonella* spp. and temperature for both samples was set as indicator to determine the survival of *Salmonella* spp. in direct and non-direct inoculums. This study gives the figures of die-off rate for *Salmonella* spp. during composting. The differentiation between direct contact (Sample A) and non-contact of *Salmonella* spp. with compost material (Sample B) during composting was also revealed. The results from laboratory scale of composting study has been showed that after 8 days (which included at least at 66°C) the numbers of *Salmonella* spp. in Sample A were below the limits in UK compost standard (known as PAS 100)(BSI, 2005) which required the compost to be free of *Salmonella* spp. Meanwhile, Sample B still gives high amount of *Salmonella* spp. in even after composting for 20 days. The Table 2 summarized the numbers of *Salmonella* spp. enumeration which clearly indicates that Sample A was reached the target required by PAS 100 compared to Sample B. This result has been shown that the contact material with compost material plays main criteria in survival of *Salmonella* spp. Thus, the major mechanism involved in viability of Salmonella spp. not only depends to a simple parameter such as temperature changes.

**TABLE 2: Summarizes results for both samples in determines the survival of *Salmonella* spp.**

| *Temperature (°C)* | *Composting times (days)* | *Sample A (cfu/gww)[a]* | *Sample B (cfu/gww)[a]* |
|---|---|---|---|
| 35 | 0 | 7.30E+09 | 8.00E+09 |
| 55 | 2 | 1.00E+04 | 4.80E+09 |
| 64 | 5 | 3.00E+02 | 1.60E+08 |
| 64 | 8 | 0 | 8.00E+07 |
| 64 | 13 | 0 | 3.00E+06 |
| 64 | 20 | 0 | 5.00E+04 |

[a]Data are mean values of three replicates

# 1    INTRODUCTION

*Salmonella* spp. is well known as pathogen indicator, in standard quality of composting. The role of *Salmonella* spp. in defining the ecological quality composts products was established in the European Commission Decisions. The limit for *Salmonella* spp. densities (Absence in 50 g) is fixed in order to assign a seal quality in Commission Decision 2001/688/EC; Commission Decision 2005/384/EC (Briancesco, 2008). Meanwhile, the UK composting standard (BSI, 2005) requires composts that are sold to be free of Salmonella spp. The detection of *Salmonella* spp. is very useful in stabilization of compost material. It is related to human pathogen and very important in order to evaluate the sanitary quality of stable compost to limit the health risk (Briancesco, 2008).

In composting, inactivation of pathogen during the composting period can be effected by a number of mechanisms (Pereira-Neto et al., 1986, Wilkinson, 2008). Compost temperatures have been said as is an important factor due to its influence on the activity of microorganisms such as microbial metabolic rate and population structure (Hassen et al., 2001). Typical process of composting begins at the ambient temperatures and with a microbial community resident in the original organic material. High microbial activities will leads to a rapid increase in temperature and a period of elevated temperatures. During thermophilic stage, many of the non-thermo tolerant organisms are activated, including several pathogens. This will be followed by a gradually decrease in microbial activity, and a cooling and maturation of the composting mass (Steger et al., 2007).

According to Stentiford (1996), temperatures above 55 °C are important to maximise sanitisation. Meanwhile, temperature between 45 and 55 °C are to improve the degradation rate and between 35 and 40 °C to increase microbial diversity. The optimum temperature for composting process is considered to be approximately 60 °C according to maximizing respiration rate and $CO_2$ evolution rate. Furthermore, in thermophilic range, the optimum temperature was determined to be 54 ºC by studying the effect on oxygen uptake rate, specific growth rate and enzymatic activity of microorganisms (Miyatake and Iwabuchi, 2006). Wilkinson, (2008) described the function of temperature and the length of exposure as the most important in the inactivation of pathogen. Mostly in small scale of composting temperature is always refers as indicators for compost maturity because it relatively easy to measure during composting. According to Gajalakshmi (2005) criteria used in the evaluation of the composting process, compost stability (maturity), and quality are based on the physical and chemical characteristics of the organic material. These parameters include a drop in temperature, degree of self-heating capacity, oxygen consumption, cation-exchange capacity, organic matter, nutrient contents, and C: N ratio (Tiquia et al., 1996). Therefore, this study was designed to investigate the relationship of temperature and contact material that may also involve in pathogen activation specifically to *Salmonella* spp. The exposure to an average temperature during composting of about 55-60°C was kept for at least 3 days as it was reported sufficiently kills the vast majority of enteric pathogen (Deportes et al., 1995). The amount of *Salmonella* spp. and temperature for both samples was set as indicator to determine the survival of *Salmonella* spp. in direct and non-direct inoculums.

# 2    Material and methods

## 2.1    *Salmonella* spp. cell culture

*Salmonella enteritidis* culti-loops, ATTC 13076 (Remel, Lenexa, KS) were suspended in tryptone soya broth (Oxoid). Tryptone soya broth (TSB) was prepared earlier by suspending 30g in 1 litre of distilled water in a clean flask. The flask containing the broth was then sterilised by autoclaving at 121°C for 15 min. The cell was grown in tryptone soya broth approximately after 24 hours in fridge. The cell culture was enumerated by plating appropriate serial dilutions using sterile ringer solution on tryptone soya agar (TSA) from Oxoid. Tryptone soya agar was prepared by suspending 40g of tryptone soya agar in 1 litre of purified water and bringing it to the boil until it dissolved completely and it was then sterilized at 121°C for 15 minutes. The cells grown in the tryptone soya broth (TSB) were counted by plating out on TSA immediately after incubation for 24 hours at 37°C.

## 2.3    Laboratory-scale composting

The laboratory composting apparatus used was the same as that used in the DR4 biodegradability tests method devised by WRc (Waste Research Centre) in the United Kingdom. The material (compost mixture) was incubated under aerobic conditions in a reactor using forced aeration. The composting reactor was cylindrical (22 cm height x 8 cm diameter) with a perforated plate at the bottom to distribute the air supplied, and was loaded with 400 g of the mixture. The air was supplied using a pump at a constant flow rate of 0.5L/min, which was measured and controlled using a flow meter (tube-and-float type). In this sealed reactor, the air was forced up through the compost material, and passed out of the reactor. Then the air was passed through a condenser, to remove surplus liquid before exhausting to atmosphere. The temperature of the compost material was monitored with a thermocouple which was inserted in the sample.

### 2.4    Raw material and inocula preparation

Matured compost produced from kitchen waste was used for both types of samples in DR4 reactors. The compost mixture was shredded to an average size of 5 – 20 mm before it was placed in a laboratory compost reactor. First sample was prepared by adding appropriate amount of culture of *Salmonella* spp. solution and matured compost from kitchen waste in the ratio 6:4 (w/w). The concentration of *Salmonella* spp inoculates directly into compost material with approximately $8.0 \times 10^9$ cfu $g^{-1}$. This sample was labelled as Sample A. On the other hands, the Sample B was prepared with inoculums solution was placed in a small vial. This 10 ml vial of culture of *Salmonella* spp. solution was then placed in a middle of reactor with contains approximately $8.0 \times 10^9$ cful/ml This small vial was cap closed tightly to make sure there is no contact between *Salmonella* spp. solution and compost material.

### 2.5    Enumeration of *Salmonella* spp.

The conventional method that is used for the enumeration of *Salmonella* spp. is serial dilution followed by standard membrane filtration as recommended in the compost quality standard method PAS 100 (BSI, 2005) and the British Standard (BS EN ISO 6579:2002). A 25g subsample of compost is taken, placed into a sterile stomacher bag together with 225ml of sterile phosphate buffered saline solution and stomached for 60 seconds. The resulting suspension is then serially diluted and an aliquot of each dilution is filtered through a 0.45um cellulose nitrate filter. Each filter is then placed onto an absorbent pad onto which has been placed 1.5ml of resuscitation medium (tetrathionate broth). The pads are then incubated for 18-24 hours at 37°C. Each filter is then transferred to a sterile Petri dish containing Rambach agar and incubated for a further 24 hours at 37°C after which the resulting colonies are counted.

### 2.6    Analysis of compost characteristics

A number of different characteristics of the compost samples were also determined. The pH values were determined from a compost/water suspension extract according to the procedures described by Hu et al., (2009). The moisture content of the compost material was determined by drying a 20g sample at 105°C for 24h hours. The drying, cooling and weighting process were repeated until no change in weight was observed.

## 3    RESULTS AND DISCUSSION

### 3.1    Characteristics of the raw material

The characteristics of the matured compost material produced from kitchen waste material were investigated to produce the compost mixture (as shown in Table 1). The compost material had moisture content approximately about 40% before adding inoculums of *Salmonella* spp. solution. According to Haug (1980), the optimum moisture content recommended values are in the 50% to 70%. The moisture content of the waste mixture was maintained in the 50% to 70% range during the whole of the composting process. The moisture content is important to provide a medium for the transport of dissolved nutrients required for the metabolic and physiological activities of microorganisms (Miller, 1989, Stentiford, 1996). The initial moisture content for Sample A after inoculums with *Salmonella* spp. solution gives 57%. Meanwhile for Sample B, the suitable moisture content was prepared by adding appropriate amount of distill water until it gives approximately 55 % of moisture content.

### 3.2    Characteristics of composts

Figure 1 shows the temperature profiles of the laboratory-scale composting test. The temperature was increased from ambient to the thermophilic range over a 2 day period. It was then kept at that temperature until day 8. According to Deportes et al., (1995) temperatures of approximately 55-60 °C for at least 3 days are recommended to be effective against *Salmonella* spp. The temperature profiles in the laboratory scale composters were set to at least exceed this level of exposure as can be seen from Figure 1. The pH values were around neutral at the beginning of the composting process, and then increased as shown in Figure 1 reaching 9.0 at the end of the composting trial. This increase of pH is characteristic of many composting operations due to ammonia produce from protein decomposition (de Bertoldi et al., 1988, Haug, 1980). The moisture content in the composting process is affected by the heat generated in the mass and the amount of aeration used, these also affect the temperature and the rate of decomposition (Turner, 2002). In this study, it was found that the moisture content of the final compost was in the range 15-20% which absolutely lower than that observed in the initial mixtures.

**TABLE 1: Characteristic of compost material used in both Reactor A and Reactor B**[a]

| *Sample* | *Sample A (Direct inoculums)* | *Sample B (Inoculums in vial)* |
|---|---|---|
| pH | 5.5 | 5.7 |
| Raw Moisture Content | 40 | 40 |
| Initial Moisture Content | 57 | 55 |
| Initial *Salmonella* spp. concentrations | $8.0 \times 10^9$ cfu g$^{-1}$ | $8.0 \times 10^9$ cfu/ml |

[a]Data are mean values of three replicates

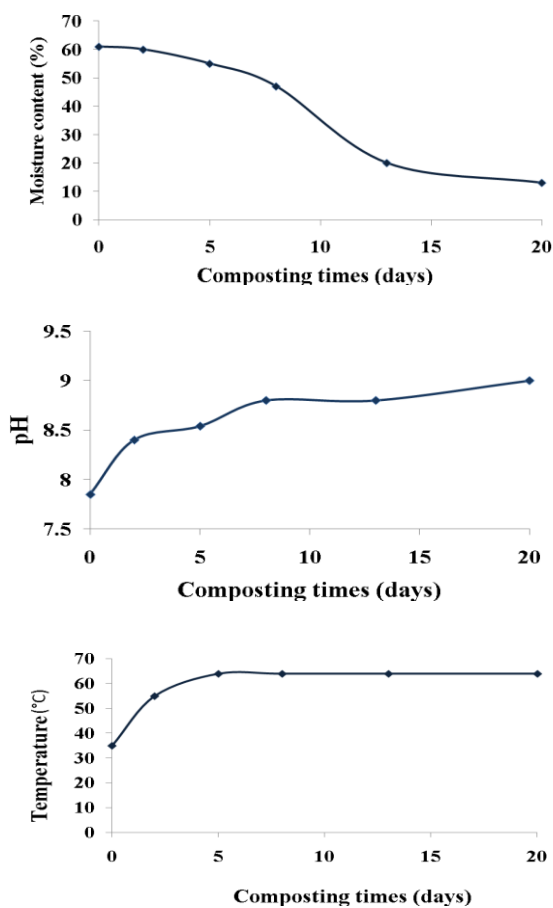

**FIGURE. 1: Moisture content, pH changes and temperature profiles of the laboratory-scale of composting test**

### 3.3 Enumerations of *Salmonella* spp. in different type of inoculums techniques

The Figure 2 shows the changes of number in *Salmonella* spp. enumeration when enumerated using the membrane filtration methods. This laboratory scale study has been shown that after composting for 8 days (which included at least at 66°C) the numbers of *Salmonella* spp. in Sample A were below the limits in UK compost standard (PAS 100) which required the compost to be free of *Salmonella* spp. Meanwhile, Sample B still gives high amount of *Salmonella* spp. in even after composting for 20 days. The Table 2 summarized the numbers from Figure 2 clearly indicates that Sample A was reached the target required by PAS 100 compared to Sample B.

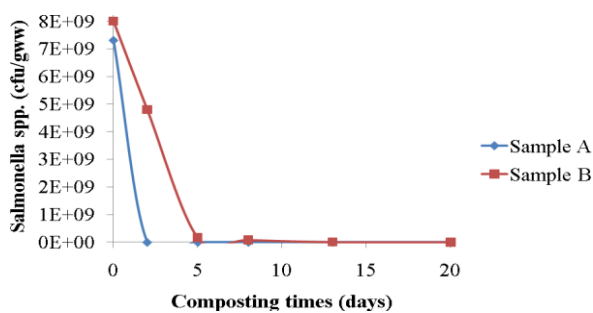

**FIGURE 2: The enumeration of *Salmonella* spp. in Sample A and B**

**TABLE 2: Summarizes results for both samples in determines the survival of *Salmonella* spp.**

| *Temperature (°C)* | *Composting times (days)* | *Sample A (cfu/gww)[a]* | *Sample B (cfu/gww)[a]* |
|---|---|---|---|
| 35 | 0 | 7.30E+09 | 8.00E+09 |
| 55 | 2 | 1.00E+04 | 4.80E+09 |
| 64 | 5 | 3.00E+02 | 1.60E+08 |
| 64 | 8 | 0 | 8.00E+07 |
| 64 | 13 | 0 | 3.00E+06 |
| 64 | 20 | 0 | 5.00E+04 |

[a]Data are mean values of three replicates

The Sample B which the *Salmonella* solution was placed into the vial was still gives approximately 5.00E+04 cfu/gww even after 20 days of composting. In this study, *Salmonella* in Sample A was gives enough time for the inactivation process. The temperature in DR4 reactor was kept at temperature 55 °C for 3 days until finally it increased to 64 °C. According to Gajalakshmi (2005), Salmonella spp. in compost material will die within 60 and 20 minutes at a temperature of 55 and 60 °C, respectively. Mostly study about inactivation of *Salmonella* spp in solution suggested that the temperature plays an important role for inactivation especially in themophilic conditions (Blackburn et al., 1997, Spinks et al., 2006). But when comparison made in this study, the contact of compost material was gives more effectively results in reducing the amount of *Salmonella* spp. According to Wilkinson, (2008) described the function of temperature and the length of exposure as the most important in the inactivation of pathogen. Therefore, the temperature was always used as main indicator to assessing the pathogen kills in composting (Miyatake and Iwabuchi, 2006). It was a relatively easy parameter to measure during the composting. This study proved that the temperature is not only the main mechanism that affected the inactivation of *Salmonella* in compost material. Even though, the both samples were set up to have a same amount of *Salmonella* spp. and temperature exposure but the direct inoculum was successfully reduced within 5 days in laboratory scale of composting. Thus, the process and contact material within compost material and *Salmonella* spp. is the major contribution for inactivation. The destruction of pathogen during the composting period can be effected by a number of mechanisms such as competition for nutrients and microbial antagonism (including antibiotic production and direct parasitism), production of organic acids and ammonia (Pereira-Neto et al., 1986, Wilkinson, 2008). Thus, it's not only depending to exposure to heat and temperature/time exposure.

## 4   CONCLUSION AND RECOMMENDATIONS

This study gives the figures of die-off rate for *Salmonella* spp. during composting. The differentiation between direct contact and non-contact of *Salmonella* spp. with compost material during composting was also revealed. The results have been showed that contact material with compost material plays main criteria in survival of *Salmonella* spp. It was a major mechanism involved in assessing pathogen killing not only depends to a typically simple parameter such as temperature changes.

## 5   ACKNOWLEDGMENT

The authors wish to thank Public Health Laboratory staff, School of Civil Engineering, University of Leeds for their valuable support and excellent laboratory facilities. This study was funded by Ministry of Higher Education of Malaysia and University of Tun Hussien Onn Malaysia.